\documentclass[12pt]{iopart}
\usepackage{graphicx}
\begin{document}

\title[Kinetic regimes in aggregating systems with fragmentation]{Kinetic regimes in aggregating systems with spontaneous and collisional fragmentation}

\author{Anna S. Bodrova}
\address{Moscow Institute of Electronics and Mathematics, National Research University Higher School of Economics, 123458, Moscow, Russia}
\address{Faculty of Physics, M. V. Lomonosov Moscow State University, Moscow, 119991, Russia}
\author{Vladimir Stadnichuk}
\address{Faculty of Physics, M. V. Lomonosov Moscow State University, Moscow, 119991, Russia}
\author{P. L. Krapivsky}
\address{Department of Physics, Boston University, Boston, MA 02215, USA}
\author{J\"urgen Schmidt}
\address{Astronomy Research Unit, University of Oulu, PL 3000 FI-90914, Finland}
\author{Nikolai V. Brilliantov}
\address{Skolkovo Institute of Science and Technology, 121205 Moscow, Russia and \\ Department of Mathematics, University of Leicester, Leicester LE1 7RH, United Kingdom}
\date{\today}

\begin{abstract}
We analyze systems of clusters and interacting upon colliding---a collision between two clusters may lead to
merging or fragmentation---and we also investigate the influence of additional spontaneous
fragmentation events. We consider both closed systems in which the total mass remains constant and open systems driven by a
source of small-mass clusters. In closed systems, the size distribution of aggregates approaches a steady state. For
these systems the relaxation time and the steady state distribution are determined mostly by spontaneous fragmentation
while collisional fragmentation plays a minor role. For open systems, in contrast, the collisional fragmentation
dominates. In this case, the system relaxes to a quasi-stationary state where cluster densities linearly grow with
time, while the functional form of the cluster size distribution persists and coincides with the steady state size
distribution of a system which has the same aggregation and fragmentation rates and only collisional fragmentation.
\end{abstract}

\pacs{81.05.Rm, 05.20.Dd, 05.40.2a}

\maketitle

\section{Introduction}

Aggregation is an important process that takes place in numerous systems and on a large variety of spatial scales \cite{Leyvraz,Krapivsky}. In everyday life it is observed when e.g. small fat globules in milk coalesce to form a cream, or in the blood clotting. Aggregation is abundant in atmospheric processes, e.g. particles of smog or other
airborne particles stick together due to the van der Waals forces \cite{agg-rev,cloud,Friedlander,Srivastava1982,BrilNatCom}. Further examples are the polymerization in solutions \cite{poly,poly2}, coagulation in colloids
\cite{col}, red blood cell aggregation \cite{cell}, aggregation of prions causing Alzheimer-like diseases
\cite{Prion}, etc. Aggregation is common in living systems such as colonies of viruses \cite{CoagVirus} or schools
of fish \cite{CoagFish}; in social systems, like internet communities \cite{Krapivsky,Dorogov,net}; in economic
networks \cite{CoagNetw}. On astronomic scales aggregation plays an important role in planetary rings \cite{PNAS}, in the coalescence of particles in interstellar dust clouds and in the formation of clusters of galaxies \cite{Ossenkopf1993}.

Aggregation is often counter-balanced by fragmentation
\cite{PNAS,MeakinErnst,FamilyMeakinErnst,MatweevPRL,BrilliantovPRE,KrapivskyPRE,colm2018}. Fragmentation may be
spontaneous, e.g. caused by thermal fluctuations like in polymer solutions \cite{poly,poly2}; it may be also of mechanical origin, like shattering of particles in planetary rings due to meteoroid bombardment \cite{Esposito, Tiscareno}. Aggregates can also break when they collide. This is believed to be an important process in Saturn's dense rings, shaping the size distribution of clusters of ring particles as a subtle balance between aggregation and fragmentation \cite{PNAS,Esposito,zebker,Cuzzi,volodya}. The breakage of particles may
be also induced by external forces \cite{ChengRedner}.

A kinetic theory \cite{PNAS,statphys} that takes into account both aggregation and fragmentation occurring when aggregates collide, relies on Smoluchowski-like equations for densities of various cluster species. In applications, there are usually clusters of minimal mass (monomers)  which cannot be split into smaller objects and heavier clusters are composed of monomers \cite{Leyvraz,Krapivsky}. Depending on the system, the physical nature of the monomers may be very different, ranging from functional chemical groups, which can associate into larger molecules, to icy particles forming size-polydisperse agglomerates in Saturn's rings. In Ref.~\cite{PNAS} this framework was applied to the size distribution of particles in Saturn's rings leading to a good agreement with observations \cite{zebker}.

There are many possible generalizations of the setting studied in \cite{PNAS,statphys} and some of them are studied in this paper. Specifically, we explore the role of spontaneous fragmentation and the effect of a source of particles of small mass. The precise nature of the source plays a negligible role and we focus on the simplest case when monomers are injected uniformly into the system. The monomers may be of very different nature, e.g. proteins in biological applications or micron-sized ice particles in the plumes of Saturn's moon Enceladus \cite{Gao2016,nature1,nature2}.

Open aggregating systems driven by input have been studied in the past \cite{hisao,kl,lu1,
lu2,colm2011,colm2012,colm2017}. In applications, however, different aggregation and fragmentation mechanisms may be
present simultaneously and it is interesting to investigate the competition of these processes. Moreover, the role of a
source  term in the evolution kinetics in such systems and its impact on the particle size distribution has not been
analyzed. In the present study we address this problem theoretically and numerically. We observe that in the absence of
a monomer source, the process of spontaneous fragmentation plays a dominant role. In contrast, collisional
fragmentation dominates if a monomer source is present. In the latter case, the systems approach a quasi steady-state
cluster size distribution where the densities evolve in a self-similar manner, keeping the shape of the size
distribution unchanged. Interestingly, the form of this size distribution corresponds to the steady-state distribution
of a system where only collisional fragmentation is present.

The remainder of the paper is organized as follows. In Section 2 we present the basic kinetic equations. In Section 3
we consider different models for the kinetic coefficients, characterizing aggregation and fragmentation rates.  In
Section 4 we analyze closed systems, while in Section 5 we study open systems driven by a source of monomers. In
Section 6 we summarize our findings.

\section{Aggregation-fragmentation equations}

We assume that all aggregates are composed of an integer number of  monomers of mass $m_1=1$. Hence $m_k=km_1=k$ is
the mass of an aggregate comprised of $k$ monomers. We thus tacitly assume that each aggregate is parametrized by one number, its mass. Furthermore, we shall consider only spatially uniform systems.

The aggregates grow by sticking together in the aggregation process that  may be symbolically written  as
$$
[i] +[j]  \longrightarrow  [i+j] \,.
$$
Let $K_{ij}$ be the rate at which this happens, quantifying the number of aggregates of size $(i+j)$ that
appear during a unit time in a unit volume from merging of aggregates of mass $i$ and $j$. In this notation the kinetic
equations, describing the time evolution of the number the densities $n_k$ of aggregates of mass $k$, can be written as
\begin{equation}
\label{eq:1} \frac{dn_k}{dt} = \frac{1}{2} \sum_{i+j=k}K_{ij}n_in_j - \sum_{i=1}^{\infty}K_{ki}n_in_k
\;\;\;\;\;\;\;\;\;\;  k=1, 2, \ldots
\end{equation}
These are the standard Smoluchowski equations \cite{smol1,smol2,C43}. The first term in the above equation describes the rate at which aggregates of size $k$ are formed from particles  of mass $i$ and $j$. The summation extends over all $i\geq 1$ and $j\geq 1$ with $i+j=k$, and the factor $\frac12$ prevents double counting. The second term gives the rate at which particles of mass $k$ disappear through merging. As the system evolves, larger and larger aggregates emerge, so mathematically there are infinitely many coupled ordinary differential equations.

In fragmentation, a cluster splits into smaller clusters. Spontaneous fragmentation is represented by the reaction scheme
\begin{equation}
\label{spont-fragm}
[k ] \longrightarrow [i_1] +[i_2] + \ldots + [i_l] \,
\end{equation}
with $i_1 +i_2+ \ldots i_l =k$ due to mass conservation.

Fragmentation may be also triggered by collisions of aggregates if the kinetic energy of their relative motion
exceeds a cerrtain threshold;  this energy transforms then into kinetic energy and surface energy of the debris. In the present study we limit ourselves to binary collisions and the collision-induced fragmentation process in this situation 
may be symbolically written as
\begin{equation}
\label{coll-fragm}
[k]+[j] \longrightarrow [i_1] +[i_2] + \ldots + [i_l] \,
\end{equation}
where again $i_1 +i_2+ \ldots + i_l =k+j$ due to mass conservation.  The possible number of outgoing clusters may vary with the masses of clusters. In most analyses, see e.g. \cite{MeakinErnst,FamilyMeakinErnst,statphys}, the number of outgoing clusters was assumed to be minimal, $l=2$ in the process \eref{spont-fragm} and $l=3$ in the process \eref{coll-fragm}. Here we study another extreme with maximal number of outgoing clusters, e.g. $l=j+k$ in the process \eref{coll-fragm}. This model postulating that in a disruptive collision particles break completely into monomers, $i_1=i_2=\ldots = i_l=1$, is extreme but it leads to essentially the same size distribution as in a class of models with sufficiently steep power-law distribution of fragment masses, and more generally for models in which small mass debris dominates \cite{PNAS}.

Whenever we consider open systems with a source of monomers, we assume that the input rate $J$ is constant and the (spatially uniform) source is turned on at $t=0$. The kinetic equations describing the processes of aggregation with
the rates $K_{ij}$ and spontaneous and collisional fragmentation with the  rates $F_k$ and $F_{ij}$, respectively, read
\begin{eqnarray}
\label{systemfull}
\fl\frac{d n_1}{dt} &=& J  - n_1\sum_{j\geq 1}K_{1j}n_j + n_1\sum_{j\geq 2}jF_{1j}n_j+\frac{1}{2} \sum_{i,j\geq 2}F_{ij}(i+j) n_in_j  + \sum_{j\geq 2}jF_j n_j \,, \nonumber\\
\fl\frac{dn_k}{dt} &=& \frac12 \sum_{i+j=k}  K_{ij} n_in_j - \sum_{i\geq 1} \left(K_{ik}+F_{ik}\right) n_in_k-F_k n_k\,\,\,,\;\;\; k\ge 2.
\end{eqnarray}
The first equation describes the evolution of the monomer density $n_1$, while the second equation accounts for the evolution of densities $n_k$ of aggregates of mass $k\ge 2$. Hereinafter we assume that both  spontaneous and collisional fragmentation processes are complete.

\section{Models for the kinetic coefficients}

There are different models for the rates of aggregation and fragmentation  depending on the particular type of motion of aggregates in the system. For polymeric and colloidal solutions particles move diffusively between collisions. In this case the merging rates are $K_{ij} = 2\pi (\sigma_i+\sigma_j)(D_i+D_j)$ in three dimensions, where $\sigma_{i}=\sigma_1 i^{1/3}$ is the diameter and $D_i$ the diffusion coefficient. The diffusion coefficient for a particle of diameter $\sigma$ is $D= B/\sigma$, where $B$ is a constant that depends on the properties of the solution. Hence 
\begin{eqnarray}
\label{eq:diff1}
K_{ij} =(i^{1/3} +j^{1/3}) (i^{-1/3} +j^{-1/3}) =2 +(i/j)^{1/3} + (j/i)^{1/3}.
\end{eqnarray}
(Hereinafter the amplitudes are set to unity; this can be done e.g. by changing the units of time.) 
The pure aggregation model with this Brownian kernel has not been solved. Smoluchowski noticed \cite{smol1,smol2,C43} the homogeneity property, $K(ai,aj)=K(i,j)$, of the Brownian kernel and suggested to consider a simpler model with constant kinetic coefficients that has the same homogeneity property. The model with $K_{ij}=\rm{const}$ is analytically tractable, it helped to develop scaling approaches which also apply to more complicated aggregation processes \cite{Leyvraz,Krapivsky}. The model with constant kernel, as well as its solvable cousins with sum and product kernels, $K_{ij} = i+j$ and $K_{ij} = ij$, played a role similar to the role of the Ising model in studies of phase transitions.

In a collisional planetary ring the particles move freely on ballistic trajectories between binary collisions. For such a ballistic aggregation the reaction rates depend on the velocity dispersions of  colliding particles and the cross-sections $\sigma_{ij}$
\begin{equation*}
K_{ij}\sim \sigma_{ij}^2 \sqrt{\langle v_i^2 \rangle + \langle v_j^2 \rangle} \\
\end{equation*}
Assuming equipartition of kinetic energies in the system,  the aggregation rates take the form of the generalized
ballistic kernel ~\cite{statphys,PNAS,Palaniswaamy2006}
\begin{equation}
\label{eq:ballist}
K_{ij} = \left(i^{1/3} + j^{1/3} \right)^{2} \left(i^{-1} + j^{-1} \right)^{1/2}
\end{equation}
Owing to dissipative collisions, in planetary rings the equipartition of kinetic energies does not hold and the  velocity dispersions of particles of different size are not very different \cite{Salo}; this motivates the following form of the kinetic coefficients
\cite{PNAS}:
\begin{equation}
\label{jur}
K_{ij} = \left(i^{1/3} + j^{1/3} \right)^{2}
\end{equation}
In the general case of a mixture of granular particles the temperature  often depends on the mass of particles
according to a  power law: $T\sim k^{\alpha}$  \cite{lev}. In this case the kinetic coefficients depend on the size of
particles in a more complicated way:
\begin{equation}
\label{Cgen} K_{ij} = \left(i^{1/3} + j^{1/3} \right)^{2} \left(i^{\alpha-1} + j^{\alpha-1} \right)^{1/2} \, .
\end{equation}
Here $\alpha=0$ corresponds to the case of equipartition, while $\alpha=1$ corresponds to the case of equal velocity dispersions of all
species.  The kernel (\ref{Cgen}) is homogeneous. For simplicity we replace this kernels with the simplified kernel of the same degree of homogeneity
\begin{equation}
\label{eq:ballist1} 
K_{ij} = \left(i j \right)^{\mu}
\end{equation}
where $\mu=1/3+(\alpha-1)/4$, which yields $\mu =1/12$ and $\mu=1/3$ for the discussed above cases of energy
equipartition and equal  velocity  dispersion. 

We study the case $\mu <1/2$ to exclude gelation. In the case of pure coagulation without fragmentation, gelation occurs \cite{Leyvraz,Krapivsky} when $\mu >1/2$. Although the gelation has not been proved for systems with fragmentation, we consider models with $\mu <1/2$ to be on the safe side. We also assume that the collisional and aggregation coefficients are proportional to each other,
\begin{equation}
\label{Aij}
F_{ij}=\lambda K_{ij}\,,
\end{equation}
see \cite{PNAS} for the justification of Eq. (\ref{Aij}). Further, we use homogeneous rates 
\begin{equation}
\label{eq:Ak}
F_k = \nu k^{\theta}
\end{equation}
for spontaneous fragmentation. The exponent $\theta$ depends on the details of the fragmentation mechanism. For instance, $\theta =0$
corresponds to the simplified model of a constant spontaneous fragmentation rate, which does not depend on the aggregate
size; $\theta =2/3$ refers to the case when the fragmentation rate is proportional to the cross-section of the
aggregate, which may happen when particles in planetary rings are disrupted by impacts of interplanetary meteoroids; $\theta=1$ mimics fragmentation of a linear aggregate whose instability is proportional to its length.

\section{Aggregation and fragmentation without injection of monomers}
\subsection{Constant rate coefficients}

Let us first consider the kinetic equations in the closed system ($J=0$) with constant kinetic coefficients, $K_{ij}=1$, $F_{ij}=\lambda$ and $F_k=\nu$. We have
\begin{eqnarray}
\nonumber
\fl\frac{dn_1}{dt}= - n_1\sum_{j\geq 1}n_j + n_1\sum_{j\geq 2}j\lambda n_j+\frac{1}{2} \sum_{i,j\geq 2}\lambda(i+j) n_in_j  + \sum_{j\geq 2}\nu j n_j \\
\fl\frac{dn_k}{dt}=\frac12\sum_{i+j=k} n_in_j - \sum_{i\geq 1} \left(1+\lambda\right) n_in_k-\nu n_k\,\,\,\;\;\;\;k\ge 2
\label{syscon}
\end{eqnarray}
Summing up all Eqs.~(\ref{syscon}) one arrives at an ordinary differential equation for the
total number density $N(t)=\sum_{i}n_i(t)$:
\begin{equation}
\label{N:eq}
\frac{dN(t)}{dt}=-N^2(t)\left(\lambda+\frac12\right)+N(t)\left(\lambda M-\nu\right)+\nu M
\end{equation}
Here $M=\sum_{k}kn_k$ is the mass density which remains constant. The number density $N(t)$ converges to the steady state solution
\begin{equation}
\label{Nbig}
N=N(\infty)=\frac{\lambda M - \nu + \eta}{1+2\lambda}
\end{equation}
where
\begin{equation}
\label{eq:eta2}
\eta=\sqrt{\left(\lambda M-\nu\right)^2+2\nu M\left(2\lambda+1\right)}\,.
\end{equation}
The solution to Eq.\ (\ref{N:eq}) reads
\begin{equation}
N(t)=\frac{N-C\left(N-\tau_{rel}^{-1}\right)e^{- t/\tau_{rel}}}{1-Ce^{-t/\tau_{rel}}}
\end{equation}
with the characteristic time
\begin{equation}
\label{eq:eta}
\tau_{rel}^{-1}=\frac{\eta}{\lambda+\frac12}\,.
\end{equation}
The constant $C$ is determined by the initial conditions
\begin{equation}
C=\frac{M-N(0)}{M-N(0)+\tau_{rel}^{-1}}
\end{equation}
The characteristic time $\tau_{rel}^{-1}$ is determined by the fragmentation coefficients $\lambda$ and $\nu$, see Eq.~(\ref{eq:eta}). If $\nu$ and $\lambda$ are small, $\nu, \, \lambda \ll 1$  and of the same
order of magnitude then $\tau_{rel}^{-1} \sim \nu^{1/2}$. In the lack of spontaneous fragmentation ($\nu=0$)
the relaxation to the steady-state occurs on a much longer timescale, $\tau_{rel}^{-1} \sim \lambda$.

All densities $n_k$ also approach a steady state. It is not possible to find the
full time-dependent solution of Eqs.~(\ref{syscon}). Nevertheless, the steady-state densities $n_k$ themselves may be found. These are the solution of the following system:
\begin{eqnarray}\nonumber
 - n_1\sum_{j\geq 1}n_j + n_1\sum_{j\geq 2}j\lambda n_j+\frac12\sum_{i,j\geq 2}\lambda(i+j) n_in_j  + \sum_{j\geq 2}\nu j n_j=0\\
\frac12\sum_{i+j=k} n_in_j - \sum_{i\geq 1} \left(1+\lambda\right) n_in_k-\nu n_k=0\,\,\,\;\;\;\;k\ge 2,
\label{syscon1}
\end{eqnarray}
which we recast into the form,
\begin{eqnarray}
\lambda MN+\nu M-n_1\left(\nu+N\left(1+\lambda\right)\right) &=&0\\
\frac12\sum_{i+j=k}n_in_j-n_k\left(\nu+\left(1+\lambda\right)N\right) &=&0\,\,\,\;\;\;\;k\ge 2.\label{musimple}
\end{eqnarray}
Multiplying Eqs.~(\ref{musimple}) by $z^k$ and performing the summation over all $k$ we get the quadratic equation for
the generating function $\mathcal{N}=\sum_{k\ge 1} n_k z^k$:
\begin{equation}
\frac12\mathcal{N}^2-\left(\nu+\left(1+\lambda\right)N\right)\mathcal{N}+\left(\nu+\left(1+\lambda\right)N\right)n_1z=0.
\end{equation}
The solution of this equation reads,
\begin{equation}
\mathcal{N}=\left(\nu+\left(1+\lambda\right)N\right)\left(1\pm\left(1-\frac{2n_1z}{\nu+\left(1+\lambda\right)N}\right)^{\frac12}\right).
\end{equation}
Using the expansion
\begin{equation}
\left(1-a\right)^{\frac12}=-\sum_{k=0}^{\infty}\frac{a^k}{k!}\frac{\Gamma\left(k-\frac12\right)}{2\sqrt{\pi}}
\end{equation}
for $a=2 n_1 z/\left(\nu+\left(1+\lambda\right)N\right)$ and the definition of the generating function $\mathcal{N}$,
we get the final expression for the number densities of particles $n_k$:
\begin{equation}
\label{anal} n_k=\frac{1}{\sqrt{4\pi}}\left(\frac{2n_1}{(1+\lambda)N+\nu}\right)^k
\left[\left(1+\lambda\right)N+\nu\right]\frac{\Gamma(k-1/2)}{\Gamma(k+1)}.
\end{equation}
Using Stirling's formula, $\Gamma (x)\simeq \sqrt{2\pi}x^{x-1/2}e^{-x}$, we finally obtain
\begin{equation}
\label{nknk} n_k=\frac{1}{\sqrt{4\pi}}k^{-3/2}\left(\frac{2n_1}{(1+\lambda)N+\nu}\right)^k
\left[\left(1+\lambda\right)N+\nu\right].
\end{equation}
The steady-state density of monomers,  $n_1$, follows from Eq.~(\ref{eq:1}),
\begin{equation}\label{n1big}
n_1=\frac{M\left(\lambda N+\nu\right)}{\left(1+\lambda\right)N+\nu},
\end{equation}
which together with Eq.~(\ref{Nbig}) for $N$ yields the final result for the densities
\begin{equation}
n_k=\frac{1}{\sqrt{4\pi}}k^{-3/2}\left(1-a\right)^k \left[\left(1+\lambda\right)N+\nu\right]
\end{equation}
where
\begin{equation}\label{aaa}
\fl
a=\frac{2\lambda^4M^2+\nu^2+2\lambda\nu\left(\nu+\eta\right)+2\nu\lambda^2\left(2M+\nu+\eta\right)+2\lambda^3M\left(2\nu+\eta\right)}{\left(\lambda^2M+\eta+\lambda\left(M+\nu+\eta\right)\right)^2}\,.
\end{equation}
Now we assume that both fragmentation constants $\lambda$ and $\nu$ are small, $ \lambda \ll 1 $ and $\nu \ll 1$.
Moreover, we assume that they are or of the same order of magnitude, $\lambda \sim \nu$, then the leading term in the
expansion of $a$ with  respect to $\lambda$ and $\nu$ reads:
\begin{equation}
a=\frac{\nu}{2M}
\end{equation}
Since $a\ll 1$, we write $\left(1-a\right)^k\simeq\exp\left(-ak\right)$ and obtain
\begin{equation}\label{nkfull}
n_k=\frac{N}{\sqrt{4\pi}}k^{-3/2} e^{-\frac{\nu}{2M} k}
\end{equation}
with $N$ is given by Eq.~(\ref{Nbig}). If we consider the case $\lambda =0$, corresponding to the absence of the binary
fragmentation, only the pre-factor $N$ in Eq. (\ref{nkfull}) will be altered,  while the expression in the exponent
will remain the same. Hence we conclude, that if both spontaneous and collisional fragmentation are of the same order of magnitude the spontaneous fragmentation always dominates and it determines the form of the steady-state aggregate size
distribution.

Consider now the case of $\nu=0$, corresponding to purely collisional fragmentation as in Ref. \cite{PNAS}. We have
\begin{equation}
n_k= (4\pi)^{-1/2}\,k^{-3/2}\left(\frac{2n_1}{(1+\lambda)N}\right)^k \left(1+\lambda\right)N
\end{equation}
with
\begin{equation}\label{LLL}
N=\frac{2M\lambda}{2\lambda+1}  \qquad \quad {\rm and} \quad \qquad n_1=\frac{\lambda M}{\left(\lambda+1\right)}.
\end{equation}
For small $\lambda\ll 1$ and large $k\gg 1$ we finally arrive at
\begin{equation}
\label{n0} n_k=\frac{M\lambda}{\sqrt{\pi}}\,k^{-3/2}\,e^{-\lambda^2 k}\,.
\end{equation}
Both dependencies, (\ref{nkfull}) and (\ref{n0}), predict a power-law size distribution with an exponential cutoff.

We note that the exponent in Eq.~(\ref{nkfull}) depends linearly on the fragmentation coefficient $\nu$, while the exponent in Eq.~(\ref{n0}) demonstrates a quadratic dependence on $\lambda$. 
This means that if $\nu$ and $\lambda$ are of comparable order of magnitude the spontaneous fragmentation will dominate. If however $\nu\ll\lambda^2$ then the collisional fragmentation dominates while spontaneous decay becomes insignificant. If $\nu\sim\lambda^2$ both fragmentation mechanisms affect the system.

\subsection{Size-dependent rate coefficients}

We turn now our attention to the case of size-dependent rate coefficients as it is given by  Eqs.~(\ref{eq:ballist1}--\ref{eq:Ak}). The steady state solution of the system of equations, Eq. (\ref{systemfull}), fulfills the conditions
\begin{eqnarray}
\nonumber
\fl- n_1\sum_{j\geq 1}j^{\mu}n_j + n_1\sum_{j\geq 2}j^{1+\mu}\lambda n_j+\frac{1}{2} \sum_{i,j\geq 2}\lambda(i+j)\left(ij\right)^{\mu} n_in_j  + \sum_{j\geq 2}\nu j^{1+\mu} n_j=0 \\
\fl \frac12\sum_{i+j=k} \left(ij\right)^{\mu} n_in_j - \sum_{i\geq 1} \left(1+\lambda\right)\left(ik\right)^{\mu} n_in_k-\nu  k^{\theta} n_k=0\,\,\,\;\;\;\;k\ge 2
\label{sysmu}
\end{eqnarray}
An analytical solution can be found for the case $\theta=\mu$. In  order to solve the system of equations we introduce
new variables
\begin{equation} \label{li}
l_k=k^{\mu}n_k.
\end{equation}
Note that $l_1=n_1$. In the steady state the system of equations (\ref{sysmu}) then reads:
\begin{eqnarray}\nonumber
 - l_1\sum_{j\geq 1}l_j + l_1\sum_{j\geq 2}j\lambda l_j+\frac12\sum_{i,j\geq 2}\lambda(i+j) l_il_j  + \sum_{j\geq 2}\nu j l_j=0\\
\frac12\sum_{i+j=k} l_il_j - \sum_{i\geq 1} \left(1+\lambda\right) l_il_k-\nu l_k = 0\,\,\,\;\;\;\;k\ge 2.\label{sysl}
\end{eqnarray}
This system has exactly the same form as Eqs.~(\ref{syscon1}), hence the solution for $l_k$ possesses the same form as
Eq. (\ref{anal}). Recalling that $n_k=l_kk^{-\mu}$ and writing $L=\sum_{i}l_i$ we obtain
\begin{equation}
\label{anal1} n_k=\frac{1}{\sqrt{4\pi}}k^{-\mu}\left(\frac{2n_1}{(1+\lambda)L+\nu}\right)^k
\left[\left(1+\lambda\right)L+\nu\right]\frac{\Gamma(k-1/2)}{\Gamma(k+1)}
\end{equation}
and, after applying the Stirling's formula,
\begin{equation}\label{nkk}
n_k=\frac{1}{\sqrt{4\pi}}k^{-3/2-\mu}\left(\frac{2n_1}{(1+\lambda)L+\nu}\right)^k
\left[\left(1+\lambda\right)L+\nu\right].
\end{equation}
Expressing $n_1=l_1$ from the first equation of the system (\ref{sysl}) and shortly writing $\tilde{M} \equiv \sum_{k}kl_k$, we have

\begin{equation}\label{n11}
n_1=\frac{\tilde{M}\left(\lambda L+\nu\right)}{\left(1+\lambda\right)L+\nu}.
\end{equation}
With Eq.~(\ref{n11}), we recast Eq.~(\ref{nkk}) into the form
\begin{equation}
\label{eq:nk} n_k=\frac{1}{\sqrt{4\pi}}k^{-3/2-\mu}\left(1-\tilde{a}\right)^k \left[\left(1+\lambda\right)L+\nu\right],
\end{equation}
where
\begin{eqnarray}
\label{eq:a1} \fl \tilde{a} =\frac{2\lambda^4\tilde{M}^2+\nu^2+2\lambda\nu\left(\nu+\tilde{\eta}\right)+
2\nu\lambda^2\left(2\tilde{M}+\nu+\tilde{\eta}\right)+2\lambda^3\tilde{M}\left(2\nu+\tilde{\eta}\right)}{\left(\lambda^2\tilde{M}+\tilde{\eta}+\lambda\left(\tilde{M}+\nu+\tilde{\eta}\right)\right)^2},
\\
\fl\tilde{\eta} = \sqrt{\left(\lambda \tilde{M}-\nu\right)^2+2\nu \tilde{M}\left(2\lambda+1\right)} . \nonumber
\end{eqnarray}
are derived similarly to (\ref{aaa}). The quantity $L$ can be expressed in terms of $\tilde{M}$ in the same way as the total
number density $N$ is expressed in terms of $M$ for the case of constant coefficients. Replacing $N$ by $L$ and $M$ by
$\tilde{M}$ in Eq.~(\ref{Nbig}), we obtain:
\begin{eqnarray}\label{Lbig}
L(\tilde{M})=\frac{\lambda \tilde{M} - \nu + \tilde{\eta}}{1+2\lambda}.
\end{eqnarray}
Using the definition of $\tilde{M}$ and Eq. (\ref{eq:nk}) for $n_k$, we write:
\begin{equation}
\label{eq:M1} \tilde{M} =\sum_{k=1}^{\infty} k^{\mu+1}n_k \simeq \frac{1}{\sqrt{4\pi}}
\left[\left(1+\lambda\right)L+\nu\right] {\rm Li}_{1/2}(1-\tilde{a}),
\end{equation}
where ${\rm Li}_{1/2}(x)$ is the polylogarithm function. Here we have assumed that terms with large $k$ have the most significant contribution. Equations (\ref{Lbig})--(\ref{eq:M1}) together with
(\ref{eq:a1}) allow us to find $\tilde{M}$ and $L$ and hence to obtain approximate expressions for the densities $n_k$ from Eq.~(\ref{eq:nk}).

When $\nu \sim \lambda \ll 1$, the leading term in the expansion of $\tilde{a}$ reads
\begin{equation}
\tilde{a}=\frac{\nu}{2\tilde{M}}\,.
\end{equation}
With $\left(1-\tilde{a}\right)^k\simeq\exp\left(-\tilde{a}k\right)$ we simplify the size distribution,
\begin{equation}
n_k=\frac{L}{\sqrt{4\pi}}k^{-3/2} e^{-\frac{\nu}{2\tilde{M}} k},
\end{equation}
where $L$ and $\tilde{M}$ are solutions of Eqs.~(\ref{Lbig}) and (\ref{eq:M1}). Approximating for $\tilde{a} \ll 1 $ the
polylogarithm function in Eq.~(\ref{eq:M1}) as ${\rm Li}_{1/2}(1-\tilde{a}) \simeq \sqrt{\pi/\tilde{a}}$ we also simplify Eq.~(\ref{eq:M1}):

\begin{equation}
\label{eq:M1sim}
\tilde{M} \simeq \frac{1}{2\sqrt{\tilde{a}}} \left[\left(1+\lambda\right)L+\nu\right] .
\end{equation}

For the case of spontaneous fragmentation, $\lambda=0$, only the prefactor of $L$ will be altered while the expression in the exponent remains the same (up to terms of second-order in the small parameters $\lambda$ and $\nu$). In this
case $\tilde{\eta}= \sqrt{\nu^2 +2 \nu \tilde{M}}$ and $\tilde{a}=\nu^2/\tilde{\eta}^2$, and Eqs.~(\ref{Lbig}) and (\ref{eq:M1}) yield,
$$
\tilde{M} \simeq  \frac{1}{2}\left(1-\frac{\nu}{4} \right), \qquad \qquad L \simeq  \sqrt{\nu} -\nu .
$$
For the case of purely collisional fragmentation, $\nu=0$, we obtain $\tilde{\eta} =\lambda \tilde{M}$ and $\tilde{a}= \lambda^2/(1+\lambda)^2
\simeq \lambda^2$. From Eq. (\ref{eq:nk}) then follows (see also \cite{PNAS}),
\begin{equation}\label{nkL}
n_k \simeq \frac{L}{\sqrt{4\pi}}k^{-3/2-\mu}e^{-\lambda^2 k}
\end{equation}
Equations (\ref{Lbig}) and (\ref{eq:M1}) lead in this case to an identity, therefore we use the relation for the total mass,
\begin{equation}\label{eq:totM}
M \simeq \int_0^{\infty}dk\, k\,n_k= \frac{L}{\sqrt{4\pi}}\frac{\Gamma\left(\frac12-\mu\right)}{\lambda^{1-2\mu}},
\end{equation}
which allows us to express $L$ in terms of the mass $M$:
\begin{equation}\label{nkL1}
n_k \simeq \frac{M \, \lambda^{1-2\mu}}{\Gamma\left( \frac12 - \mu \right) } k^{-3/2-\mu}e^{-\lambda^2 k}.
\end{equation}

\begin{figure}\centerline{\includegraphics[width=0.8\textwidth]{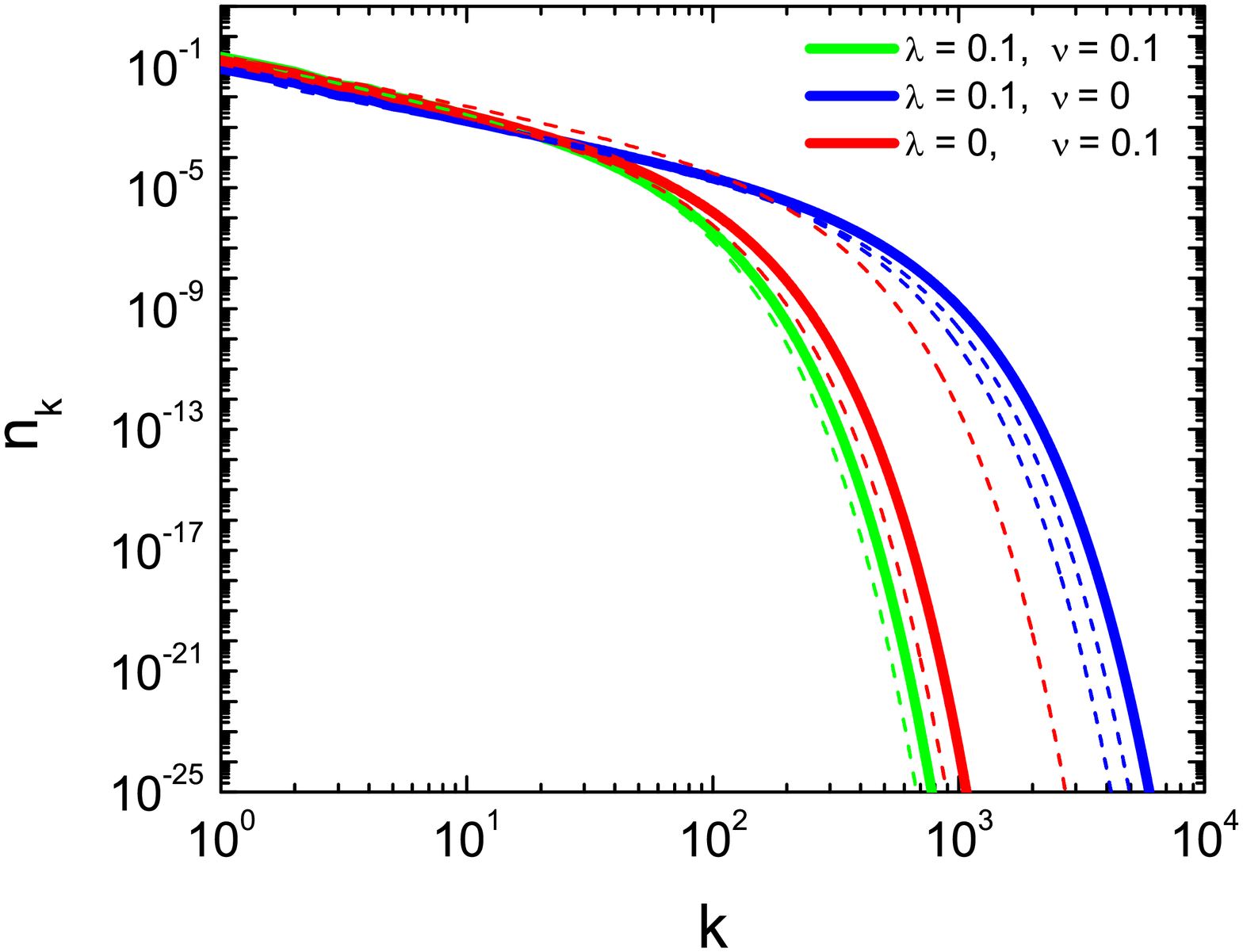}}
\caption{ Steady-state size distribution of densities $n_k$ for the rates $K_{ij}=\left(ij\right)^{\mu}$,
$F_{ij}=\lambda K_{ij}$ and  $F_k=\nu  k^{\mu}$, with $\mu=1/12$. The total number of equations is $N_{eq}=16000$. A
good agreement between the numerical (solid lines) and analytical solution [(dashed lines), Eq. (\ref{anal})] is
observed. A slight deviation from the exact solution for large $k$ may be attributed to the computational errors, when
very small numbers are handled. It can be seen that if spontaneous fragmentation is present in the system,  then the
presence (green line) or absence (red line) of collisional fragmentation does not significantly affect the system.} \label{Gmu112}
\end{figure}

In order to find the numerical solution of a very large number of rate equations, we use the fast numerical algorithm
proposed in Ref. \cite{volodya}. The numerical and analytical steady-state solutions of the system of equations
(\ref{sysmu}) are depicted at Fig.~\ref{Gmu112}. The solutions for $\nu=0.1, \lambda=0$ and $\nu=0.1, \lambda=0.1$ are
very close to each other. This illustrates that when both fragmentation mechanisms are present with comparable small
rates of the same order of magnitude, the spontaneous fragmentation dominates and determines the resulting steady-state
size distribution. Physically, this follows from the fact that the steady concentrations of the aggregates quickly
decrease with size.  But the rates for spontaneous fragmentation decrease linearly with densities, and the rates for
collisional fragmentation quadratically. In other words, the latter terms scale as $\sim n_k^2$, while the former like
$\sim n_k$ so that they dominate for large $k$.

\begin{figure}\centerline{\includegraphics[width=0.8\textwidth]{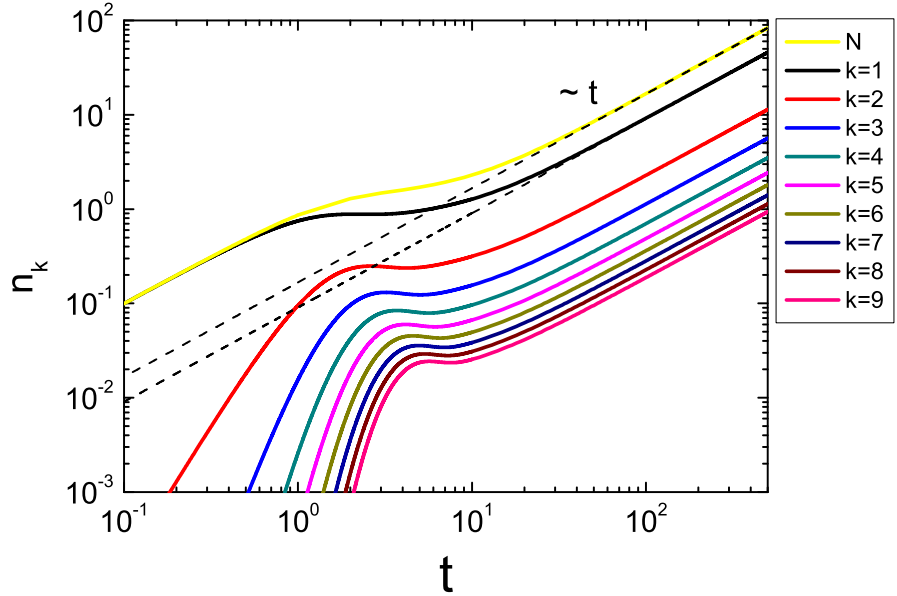}}
\caption{Evolution of the total number density $N$ and the number densities of species  $n_k$ for $k=1...9$ obtained by
the numerical solution of the system of equations (\ref{sysconst}) with $J=1$. At large time, $t \gg 1$ all
densities increase linearly with time, $n_k \sim t$.  The dashed lines correspond to fitting (bottom to top) with
$n_1/(Jt)=\lambda /\left(\lambda+1\right)$ and $N/(Jt)=2\lambda /\left(1+2\lambda\right)$.} \label{Gn1}
\end{figure}

Interestingly, the obtained steady-state size distribution, depicted in Fig. \ref{Gn1}, obeys for $k \ll \lambda^{-2} $
a power law, $n_k \sim k^{3/2 -\mu}$. Similarly, a power-law scaling of the size distribution for small $k$, was
reported in Refs. \cite{FamilyMeakinErnst,MeakinErnst}, but for a very different collision model -- whereas we studied
a complete disintegration, a breakage into two pieces was assumed in Ref. \cite{FamilyMeakinErnst,MeakinErnst}. Note,
however, that while we find here a full solution for the size distribution, only scaling exponents have been presented
in \cite{FamilyMeakinErnst,MeakinErnst}.

\begin{figure}\centerline{\includegraphics[width=0.8\textwidth]{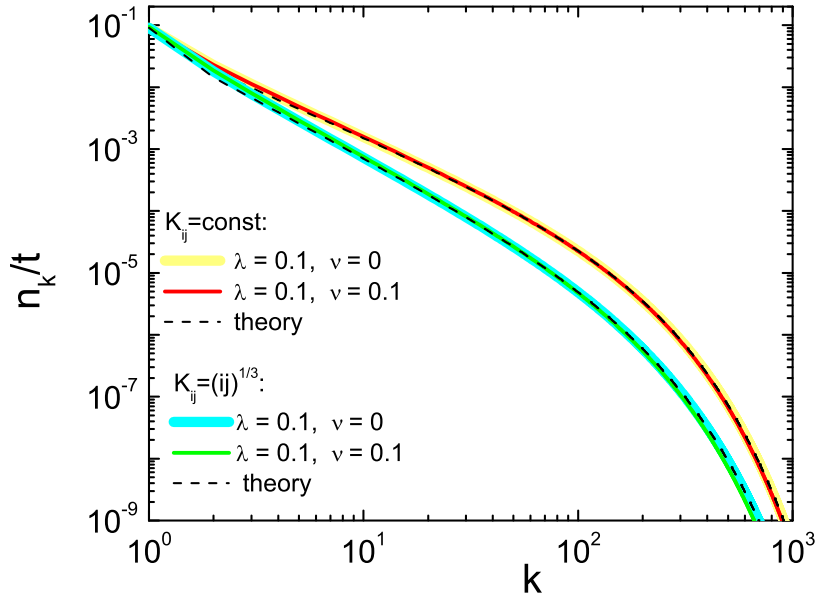}}
\caption{Quasi-stationary size distribution of particles $n_k/(Jt)$ for constant rate coefficients $K_{ij}=1$,
$F_{ij}=\lambda$, $F_k=\nu$ and size-dependent  rate coefficients $K_{ij}=\left(ij\right)^{\mu }$,
$F_{ij}=\lambda\left(ij\right)^{\mu}$ and  $F_k=\nu k^{\mu}$ for $\mu=1/3$ and $J=1$. The total number of equations is
$N_{\rm eq}=1000$, the time of evolution is $t=150$. The form of the quasi-stationary size distribution $n_k/(Jt)$
coincides with the form of a steady-state size distribution of a source-free system with the same aggregation and
fragmentation kernels for the case of collisional fragmentation only. } \label{Gconstsource}
\end{figure}

\section{Aggregation and fragmentation processes driven by a source of monomers}
\subsection{Constant rate coefficients}

First we address again the case of constant rate coefficients,  $K_{ij}=1$, $F_{ij}=\lambda$ and $F_j=\nu$.
The governing equations become
\begin{eqnarray}
\nonumber
\fl\frac{dn_1}{dt}= J- n_1\sum_{j\geq 1}n_j + n_1\sum_{j\geq 2}j\lambda n_j+\frac{1}{2} \sum_{i,j\geq 2}\lambda(i+j) n_in_j  + \sum_{j\geq 2}\nu j n_j\\
\fl\frac{dn_k}{dt}=\frac12\sum_{i+j=k} n_in_j - \sum_{i\geq 1} \left(1+\lambda\right) n_in_k-\nu  n_k\,\,\,\;\;\;\;k\ge 2
\label{sysconst}
\end{eqnarray}
With the total number density $N$ and mass $M$ the above equations may be recast into the form,
\begin{eqnarray}
\label{dn1dt}
\frac{dn_1}{dt}=\lambda MN+\nu M-n_1\left(\nu+N\left(1+\lambda\right)\right)+J  \\
\label{dnkdt}\frac{dn_k}{dt}=\frac12\sum_{i+j=k}n_in_j-n_k\left(\nu+\left(1+\lambda\right)N\right)=0 \,\,\,\;\;\;\;k\ge
2
\end{eqnarray}
Summing up all equations, we get the equation for $N$:
\begin{equation}
\label{dLdt}
\frac{dN}{dt}=-N^2\left(\lambda+\frac12\right)+N\left(\lambda M-\nu\right)+\nu M+J
\end{equation}
Naively one could expect that Eq.~(\ref{dLdt}) has the same solution as Eq.~(\ref{N:eq}). This is, however, not the case, because in Eq.~(\ref{dLdt}) the total mass is time-dependent and linearly grows with time: $M=Jt$. (For concreteness,  we assume that initially there were no particles in the system; the same asymptotic behavior emerges in the general case.)

For the total number of particles $N$ and number of monomers $n_1$ we seek solutions of the form
\begin{eqnarray}
\label{nLper1}
n_1=n_{10}t+n_{11}+n_{12}t^{-1} + \ldots \\
\label{nLper2}
 N=N_0t+N_1+N_2t^{-1}  + \ldots
\end{eqnarray}
and solve the equations perturbatively: We substitute (\ref{nLper1})--(\ref{nLper2}) into (\ref{dn1dt})--(\ref{dLdt})
and equate the coefficients at each order of $t$ separately. Keeping terms up to $O(t^{-1})$ we get
\begin{eqnarray}
\label{eq:Nfin}\fl N=\frac{2\lambda J}{1+2\lambda}\,t+\frac{\nu}{\lambda\left(1+2\lambda\right)}+\frac{2J\lambda^2-\nu^2\left(2\lambda+1\right)}{2\lambda^3J\left(1+2\lambda\right)}\,t^{-1}\\
\label{eq:n1fin}\fl n_1=\frac{\lambda
J}{\lambda+1}\,t+\frac{\left(1+2\lambda\right)\nu}{2\lambda\left(1+\lambda\right)^2}
+\left[\frac{1+2\lambda}{2\lambda\left(1+\lambda\right)^2}
-\frac{\nu^2\left(1+2\lambda\right)\left(2\lambda+2\lambda^2+1\right)}{4\lambda^3J\left(1+\lambda\right)^3}\right]t^{-1}.
\end{eqnarray}
Similar result for the simplified case of $\nu =0$ has been reported in \cite{Timokhin2019}. Taking into account that
$M=Jt$, one can see that the coefficients $N_0$ and $n_{10}$ coincide, respectively, with the total density $N$ and the
density of monomers $n_1$, for the case of purely collisional fragmentation ($\nu=0$) in a system without a monomer
source, Eqs.~(\ref{LLL}):
\begin{eqnarray}\label{LLLt}
\frac{N}{M}=\frac{N}{Jt}=\frac{2\lambda }{1+2\lambda}\\
\frac{n_1}{M}=\frac{n_1}{Jt}=\frac{\lambda }{\lambda+1}\label{n1n1t}
\end{eqnarray}
These terms do not depend on the rate of the spontaneous fragmentation $\nu$, although all particles (apart from
monomers) undergo spontaneous fragmentation. Obviously this  is a consequence of the fact, that the intensity of the
collision fragmentation, determined by the product of two concentrations, grows quadratically with time, since
concentrations grow linearly. The intensity of the spontaneous fragmentation grows, however, linearly with time as the
concentrations. Asymptotically, for $t\to \infty$, the former mechanism completely shadows the latter, which yields
$\nu$-independent $N_0$ and $n_{1 0}$.

Hence the terms containing $N_0$ and $n_{10}$ grow linearly with time and are dominant. This is confirmed by the
numerical solution of the system of rate equations (\ref{sysconst}) and illustrated in  Fig.~\ref{Gn1}. For large time
($t\gg 1$), the density of monomers $n_1$ attains the asymptotic  form given by Eq.~(\ref{n1n1t}) and the total cluster
density grows according to Eq.~(\ref{LLLt}). This asymptotic behavior is shown in Fig.~\ref{Gn1}. Generally, it is
straightforward to show that all densities $n_k$ grow linearly with time, giving rise to a \emph{quasi-stationary}
steady state $n_k/M=n_k/Jt=n_{k0}$, where $n_{k0}$ is the solution to the system of rate equations with the collisional
fragmentation only ($\nu=0$) in the absence of the monomer source, Eq. (\ref{n0}):
\begin{equation}
\label{eq:nk0} n_{k0} = \frac{n_k}{Jt}=\frac{\lambda}{\sqrt{\pi} }k^{-3/2}e^{-\lambda^2 k}
\end{equation}
The solutions (\ref{eq:nk0}), (\ref{LLLt}) and (\ref{n1n1t}) may be called quasi-stationary, since the form of the
reduced density distribution, $n_{k0}=n_k/M$, scaled with the total mass $M$, persists while all densities grow. As it follows from Eqs.~(\ref{eq:Nfin}) and  (\ref{eq:n1fin}), the relaxation to the
quasi-stationary form is completed generally, at the time $t \gg \nu/(\lambda^2 \,J)$, where all terms, except the one linear in time, may be neglected; for purely collisional fragmentation the corresponding relaxation time reads $t \gg
1/(\lambda J^{1/2})$.

By numerical solution we find that the full kinetic equations with and without spontaneous fragmentation give the same quasi-stationary size distribution, see Fig.~\ref{Gconstsource}. The influence of spontaneous fragmentation asymptotically vanishes. Qualitatively, this follows from the fact that all densities $n_k$ grow linearly with time due to permanent input of monomers and cluster aggregation. Since the collisional fragmentation depends quadratically on densities, while spontaneous only linearly, the former mechanism dominates when $t \gg  1$. This is in a sharp contrast to systems without source where spontaneous fragmentation is found to play an important role for the establishment of the cluster size distribution.

\subsection{Size-dependent rate coefficients}

For size-dependent rate coefficients the system of rate equations with the monomer source has the following form:
\begin{eqnarray}
\nonumber
\fl\frac{dn_1}{dt}= J- l_1\sum_{j\geq 1}l_j + l_1\sum_{j\geq 2}j\lambda l_j+\frac{1}{2} \sum_{i,j\geq 2}\lambda(i+j) l_il_j  + \sum_{j\geq 2}\nu j l_j\\
\fl\frac{dn_k}{dt}=\frac12\sum_{i+j=k} l_il_j - \sum_{i\geq 1} \left(1+\lambda\right) l_il_k-\nu  l_k\,\,\,\;\;\;\;k\ge 2
\end{eqnarray}
where the $l_i$ are  given by Eq.~(\ref{li}). The same analysis as for the case of constant rate coefficients leads to the
conclusion that the system behaves essentially in the same way. That is, the leading term for $l_1=n_1$ depends again linearly on time,
\begin{equation}
n_1=\frac{\lambda J}{\lambda+1}t,
\end{equation}
and the same behavior is found for other densities $n_k$. As a result, for $t \gg 1$ the system arrives also at the quasi-stationary state with $n_k/M=n_k/(Jt)$ becoming time-independent. It corresponds to the shape of the size distribution of a source-free system without  spontaneous fragmentation [see Eq.~(\ref{nkL1})]:
\begin{equation}
\frac{n_k}{Jt}=\frac{\lambda^{1-2\mu}}{\Gamma\left(\frac12-\mu\right)}k^{-3/2-\mu}e^{-\lambda^2 k}.
\end{equation}

\section{Conclusions}

Using analytical and numerical techniques we investigated a system of clusters undergoing aggregation supplemented by  collisional and spontaneous fragmentation. We analyzed both source-free systems and systems with a source of monomers. For the
aggregation rates we used a kernel that has a power-law dependence on sizes of aggregates,
$K_{ij}=(ij)^{\mu}$. For the collisional fragmentation we used the kernel $F_{ij} = \lambda K_{ij}$, where the coefficient $\lambda$ quantifies the relative frequency of the disruptive impacts. For the spontaneous fragmentation, we used a power-law kernel $F_k = \nu  k^{\theta}$. We also assumed that fragmentation is complete, that is, it results in decomposition into monomers. A physical justification of such a simplified fragmentation model has been provided in a previous study \cite{PNAS}.

We demonstrated that in source-free systems after a relaxation time, quantified by the fragmentation coefficients $\nu$ and $\lambda$, the system arrives at a steady state. Interestingly, if the fragmentation rates are of the same order of magnitude, $\lambda\sim\nu$, both the relaxation time as well as the steady state size distribution of aggregates are determined mainly by the spontaneous fragmentation, while the collisional fragmentation plays a minor role. This behavior follows from the mathematical structure of the aggregation-fragmentation equations.

Qualitatively different behaviors emerge if a source is present. A relaxation process is also observed and determined by the
fragmentation coefficients $\nu$ and $\lambda$. However, in this case the system relaxes to a quasi-stationary state in which all densities grow linearly with time, while the shape of the size distribution coincides with that of the source-free system undergoing aggregation and purely collisional fragmentation with the same rate coefficients. After the relaxation time the impact of the spontaneous fragmentation on the evolution kinetics and cluster size distribution is negligible.

The mass density in systems driven by the input of small mass clusters grows linearly with time and this often determines emerging behaviors. Different behaviors occur when the source is supplemented by the removal of large mass clusters. This setting is popular in the context of pure aggregation, but in aggregation-fragmentation systems it was analyzed in a very few studies (see \cite{Ziff85}) and deserves further analyses.

\end{document}